\documentclass[12pt]{article} 
\usepackage{rotating}
\usepackage{placeins}
\newcommand{\vtheta}{\mbox{$\vartheta$}}
\newcommand{\vphi}{\mbox{$\varphi$}}
\newcommand{\degree}{\mbox{$^\circ$}}
\newcommand{\gev}{\mbox{$\rm GeV$}}
\newcommand{\mev}{\mbox{$\rm MeV$}}
\newcommand{\gevc}{\mbox{$\rm GeV/c$}}

\newcommand{\gevcsq}{\mbox{$\rm (GeV/c)^2$}}
\newcommand{\heep}{\mbox{$^1$H(e,e$^\prime$p)}}
\newcommand{\deep}{\mbox{D(e,e$^\prime$p)n}}

\newcommand{\emiss}{\mbox{$\epsilon_m$}}

\newcommand{\qmag}{\mbox{$\vert \vec q \vert$}}

\newcommand{\ppf}{\mbox{$p_f$}}

\newcommand{\ppm}{\mbox{$p_{m}$}}

\newcommand{\einc}{\mbox{$E_i$}}

\newcommand{\qsq}{\mbox{$Q^2$}}
\newcommand{\Qsq}{\mbox{$Q^2$}}
\newcommand{\xbj}{\mbox{$x_{bj}$}}
\newcommand{\thetap}{\mbox{$\vtheta_p$}}
\newcommand{\thetapq}{\mbox{$\vtheta_{pq}$}}
\newcommand{\thetanq}{\mbox{$\vtheta_{nq}$}}
\newcommand{\phip}{\mbox{$\vphi_p$}}

\newcommand{\thetae}{\mbox{$\vtheta_e$}}
\newcommand{\phie}{\mbox{$\vphi_e$}}

\newcommand{\exx}[1]{\mbox{$\cdot 10^{#1}$}}
\begin{document}

\begin{center}
\bf{
\Large{Deuteron Electro-Disintegration at Very High Missing Momenta}
}
\end{center}

\begin{center}
K.~Aniol \\
{\em California State University L.A.} \\
\vspace*{0.25cm} 
F.~Benmokhtar\\
{\em Carnegie Mellon University} \\
\vspace*{0.25cm} 
W.U.~Boeglin (spokesperson), P.E. Markowitz, B.A.~Raue, \\
  J.~Reinhold and M.~Sargsian \\
  {\em Florida International University} \\
 \vspace*{0.25cm} 
C.~Keppel, M.~Kohl \\
{\em Hampton University}\\
 \vspace*{0.25cm} 
 D.~Gaskell, D.~Higinbotham, M.~K.~Jones (co-spokesperson),G.~Smith and 
S.~Wood \\
 {\em Jefferson Lab} \\
  \vspace*{0.25cm} 
S.~Jeschonnek\\
{\em Ohio State University} \\
  \vspace*{0.25cm} 
J.~W.~Van Orden\\
{\em Old Dominion University} \\
 \vspace*{0.25cm} 
G.~Huber\\
{\em University of Regina} \\
 \vspace*{0.25cm} 
E.~Piasetzky, G.~Ron, R.~Shneor \\
{\em Tel-Aviv University} \\
  \vspace*{0.25cm} 
H.~Bitao \\
{\em Lanzhou University} \\
  \vspace*{0.25cm} 
X. Jiang, A. Puckett\\
{\em Los Alamos National Laboratory} \\
\newpage
S.~Danagoulian \\
{\em North Carolina A\&T State University} \\
  \vspace*{0.25cm} 
H.~Baghdasaryan, D.~Day, N.~Kalantarians, R.~Subedi \\
{\em University of Virginia} \\
  \vspace*{0.25cm} 
F.~R.~Wesselmann \\
{\em Xavier University of Louisiana}\\
  \vspace*{0.25cm} 
A.~Asaturyan, A.~Mkrtchyan, H.~Mkrtchyan, V.~Tadevosyan and S.~Zhamkochyan\\
{\em Yerevan Physics Institute}

\end{center}

%
\clearpage
\begin{abstract}
  We propose to measure the $\deep$ cross section at $\qsq=4.25$
  $\gevcsq$ and $\xbj=1.35$ for missing momenta ranging from $\ppm =
  0.5$ $\gevc$ to $\ppm = 1.0$ $\gevc$ expanding the range of missing
  momenta explored in the Hall A experiment (E01-020). At these energy
  and momentum transfers, calculations based on the eikonal approximation have been shown to
   be valid and recent experiments indicated that final state
  interactions are relatively small and possibly independent of
  missing momenta. This experiment will provide for the first time
  data in this kinematic regime which are of fundamental importance to
  the study of short range correlations and high density fluctuations
  in nuclei. The proposed experiment could serve as a
  commissioning experiment of the new SHMS together with the HMS in Hall C. 
  A total beam time of 21 days is requested.
\end{abstract}
\clearpage
\section{Contribution to the Hall C Upgrade}
Werner Boeglin and Joerg Reinhold plan to contribute to the Hall C Analysis Software development.
Werner Boeglin has previously contributed to the old Hall A analyzer ESPACE and 
participated in numerous Hall C experiment as run coordinator.
Starting with the first Hall C experiments in 1995/96 Joerg Reinhold has not 
only managed numerous experiments as run coordinator and spokesperson, but also 
coordinated the corresponding analysis efforts. Thus, he has extensive 
knowledge of the existing Hall C software down to the source code level. 
He already started discussions with the SHMS-HMS User Board on how he could contribute. 
More firm tasks will be assigned during the next Hall C meeting in January 2010.

Pete Markowitz reaffirms his previously made commitment to work on the SHMS commissioning, as well 
as the software and data acquisition upgrades. The SHMS will require verification of 
the optics and measurements of acceptance and detector efficiencies. 
He has previously worked on the Halls A and C spectrometer commissioning and software for Hall A analysis. 

\clearpage
\section{Physics Motivation}

High-energy, exclusive electro-disintegration of the deuteron is
considered as the most effective process in probing two nucleon
dynamics at short space time distances. The latter condition is
essential for probing the limits of nucleonic degrees of freedom in
strong interaction dynamics.

Recently, several electro-production experiments involving $A>2$  
nuclei\cite{Kim1,Kim2,Eip1,Eip2} clearly demonstrated the possibility of 
probing high momentum components at  high $Q^2$ and $\xbj>1$ kinematics.
These experiments prepare the stage for the exploration of nuclear structure 
at short distances which is of fundamental importance in understanding the 
limits of the nucleonic picture of nuclei and the dynamics of the nuclear force at 
short distances.  However, the most basic system to study in this respect is the 
deuteron as many questions related to  two-nucleon short range 
correlations are directly related to high values of relative momenta 
in the $NN$ system.   The best way to probe large relative momenta in the $NN$ system 
is to study deuteron electro-disintegration at the same kinematics in  which 
the above mentioned $eA$ reactions are studied.

After an initial deuteron break-up experiment\cite{Ulm02} was carried out at a moderate value of 
$\Qsq=0.66$~$\gevcsq$,  two  new  experiments at Jefferson Lab\cite{HallB,HallA,Kims} for 
the first time probed deuteron break-up at large $Q^2$ kinematics.
They supported the claim that high $\Qsq$ and $\xbj>1$
are necessary conditions for using the deuteron break-up reaction as
an effective tool for  the investigation of large relative momenta in 
the $pn$ system.
This claim  
was based on the theoretical expectation that at these kinematics soft, two-body
processes are either suppressed (such as meson exchange currents and isobar contributions)
or under the control (final state interactions).

Currently no data exist that satisfy this condition for
missing momenta above $\ppm = 0.5$~$\gevc$. Data in this kinematic
range are of fundamental interest not only for the short range
structure of the deuteron itself, but also for the interpretation of
future experiments that probe the structure of short range 
correlations in heavier nuclei. 
In recent years considerable  progress has been made 
in the development of theoretical methods for the calculation of high $Q^2$
electro-disintegration of two and three nucleon systems~\cite{JW08,JW09,JW09_2,Laget,Ciofi,
Misak} extending the possibility of investigating the bound nucleon's momentum range
beyond $\ppm = 0.5$~$\gevc$.

\subsection{Research Subjects in Deuteron Electro-Disintegration}

Probing the deuteron at large relative momenta via electro-disintegration will make contributions to 
addressing several issues, each of them having a fundamental importance in nuclear physics.

\begin{itemize}
\item {\bf Reaction Dynamics:} At large internal momenta, the virtual photon interacts with a deeply bound nucleon whose interaction dynamics is largely unknown.  The research subjects include the structure of the electromagnetic current as well as modifications of nucleon form factors due to large off-shell effects.  The latter is part of the wider program of studies of modification of hadrons in the nuclear medium.

\item {\bf Final State Interaction:} In the break-up reaction, the proton and neutron undergo strong final state interactions~(FSI). The contribution of FSI can not be neglected and its understanding is an important condition for the success in probing the deuteron at small $pn$ separations. The advantage of high $Q^2$ is that one can satisfy the condition of eikonality of the final state interaction which allows one to sum the large number of partial waves in the $pn$ continuum into the elastic $pn$ scattering amplitude. The eikonal nature of FSI is characterized by a very specific angular dependence of the deuteron break-up reaction as a function of the neutron recoil angle which can be studied experimentally.

\item {\bf Deuteron Wave Function:} Finally one of the most fundamental aspects of studies of the deuteron break-up reaction is to probe the deuteron wave function at small inter-nucleon distances. This is related to the understanding of several issues such as the $NN$ potential at short distances, relativistic effects and non-nucleonic components as well as the transition from hadronic to quark-gluon degrees of freedom in the strongly bound $pn$ system.  

\end{itemize}

The above discussed research subjects are intertwined.  The advantage of the exclusive 
break-up reaction is that it provides a multitude of observables and kinematic settings that 
can be used to ultimately separate the various processes considered.
These observables include $Q^2$, $x$, recoil momentum and angular dependence of 
the break-up cross section as well as the measurement of  
asymmetries when polarization degrees of freedom are included.

\subsection{Experimental Status}
\begin{figure}[h]
  \begin{center}
    \includegraphics[width=.5\textwidth]{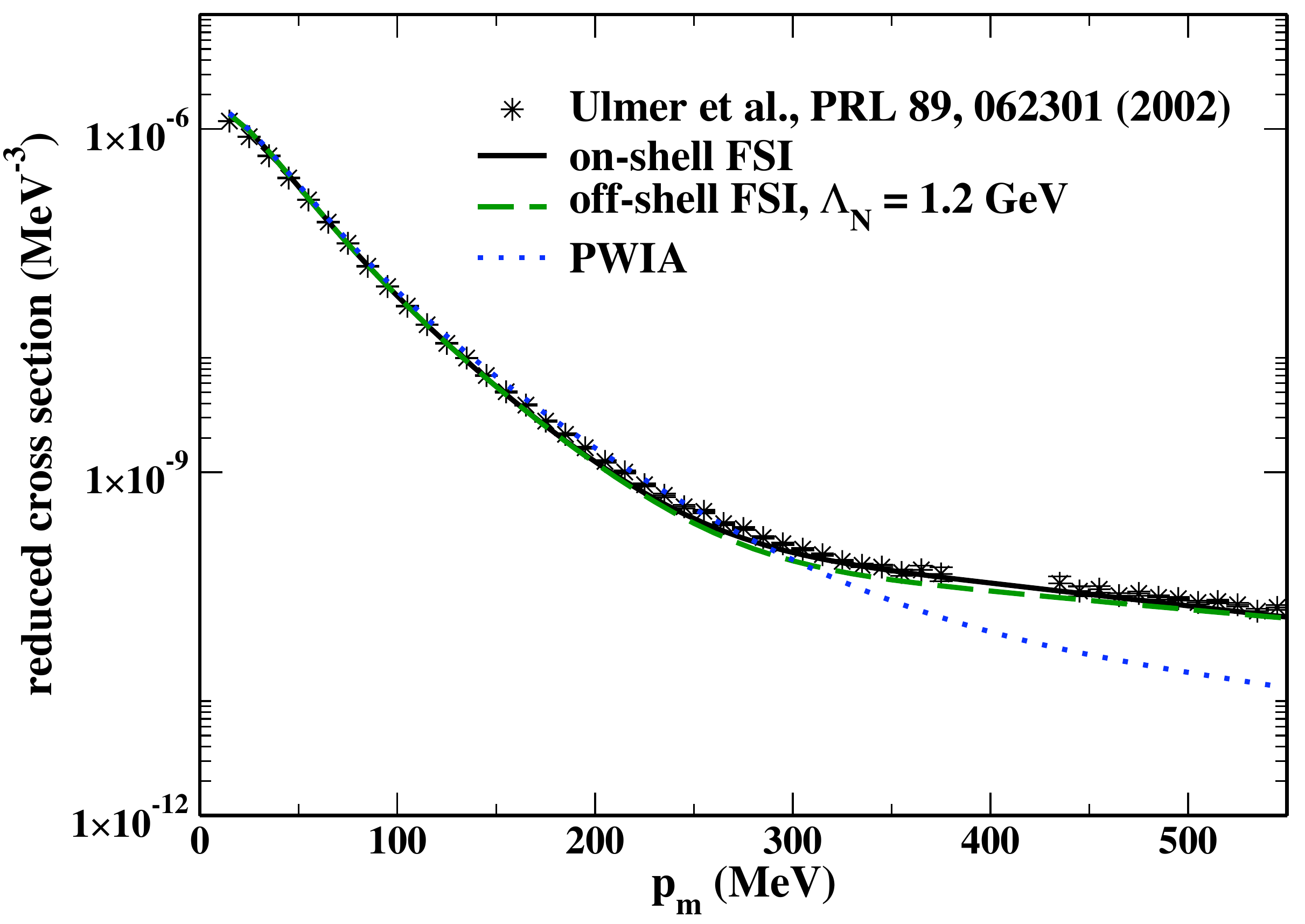}
    \caption{\footnotesize The reduced cross section for a beam energy of $3.1095$ GeV, 
$Q^2 = 0.665~\gevcsq$, $x_{Bj} = 0.964$, and $\phi_p = 180^\circ$. 
The data are from \protect{\cite{Ulm02}}.}
    \label{Fig.ulmer1}       
    \includegraphics[width=.5\textwidth]{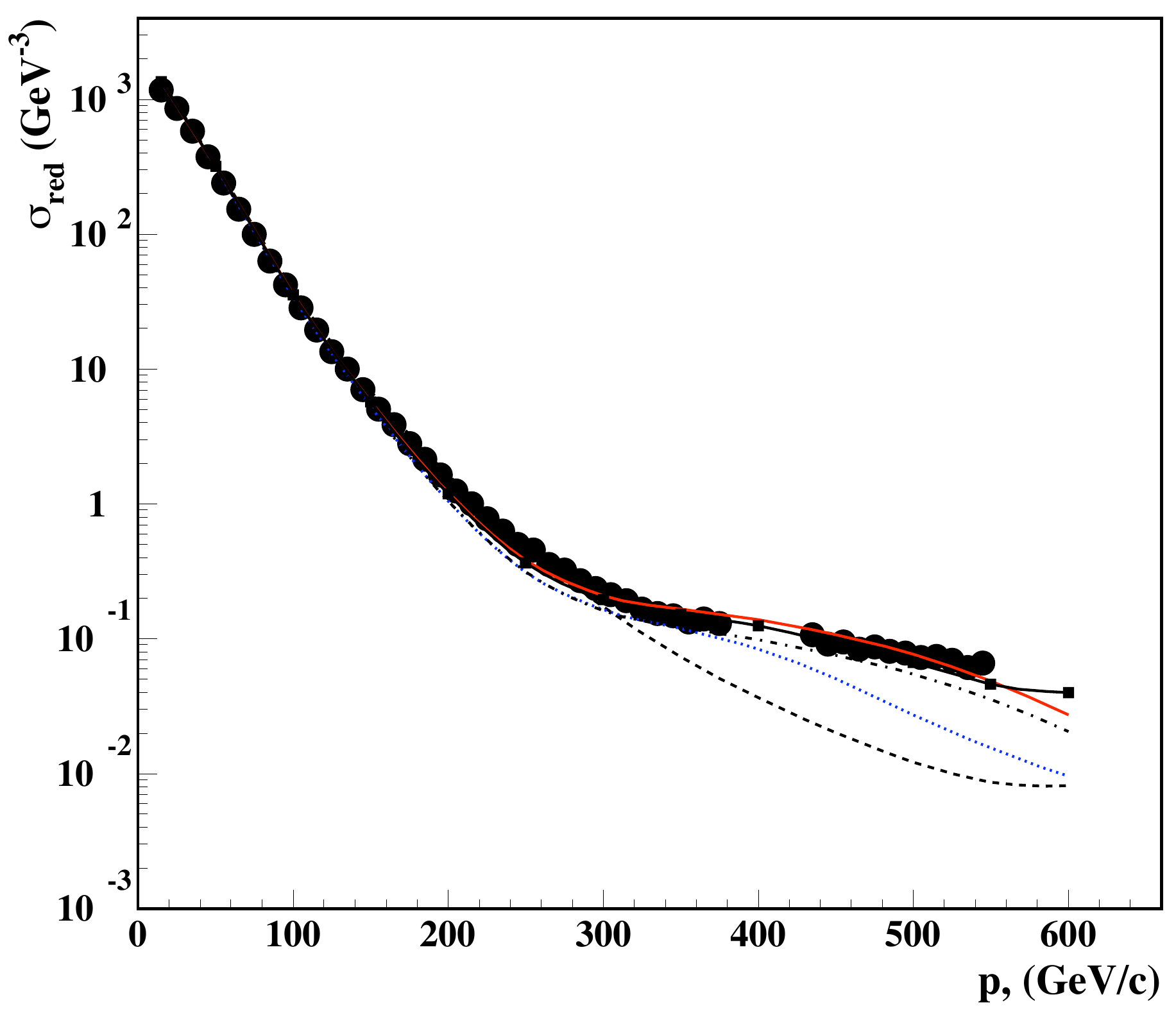}
    \caption{\footnotesize Missing momentum dependence of the reduced cross section. 
The data are from Ref.\cite{Ulm02}.
Dashed line - PWIA calculation, dotted line - PWIA+ only pole term of forward FSI,  dash-dotted line  -  PWIA+ forward FSI,  
solid line - PWIA + forward and charge exchange FSI, and solid line with squares - same as the previous solid 
line, added the  contribution from the mechanism in which the proton is a spectator and the neutron was struck  by the virtual photon}
    \label{Fig.ulmer2}       
  \end{center}
\end{figure}

A previous Hall A experiment determined the $\deep$ cross section at a
relatively low momentum transfer of $\qsq = 0.67$~$\gevcsq$ for missing
momenta up to $\ppm = 0.55$~$\gevc$ at $\xbj \approx
1$~\cite{Ulm02}. 
Even though the measured $Q^2$ was relatively small, the momentum of the final proton 
was $1$~$\gevc$ which was large enough for the eikonal approximation to be valid.
Figs.~\ref{Fig.ulmer1} and \ref{Fig.ulmer2} show the comparisons of two different 
models\cite{JW08,Misak} with the data and 
both indicate that at this kinematic setting ($x\approx 1$) final state interactions
strongly dominate the cross section especially for missing momenta above
0.4~$\gevc$.  Both calculations also show that FSI are reasonably under the control. 

This has been confirmed by two recently completed
experiments\cite{HallB,HallA,Kims} at Jefferson Lab representing the
first attempts to systematically study the exclusive deuteron break-up
reactions in the $Q^2\ge 1$~$\gevcsq$ region.  They also
confirmed that meson-exchange currents are a small correction to
the overall cross section and that isobar currents can be kept under
control by choosing $\xbj>1$\cite{Kims}.

\begin{figure}[htb]
  \begin{center}
    \includegraphics[width=.5\textwidth]{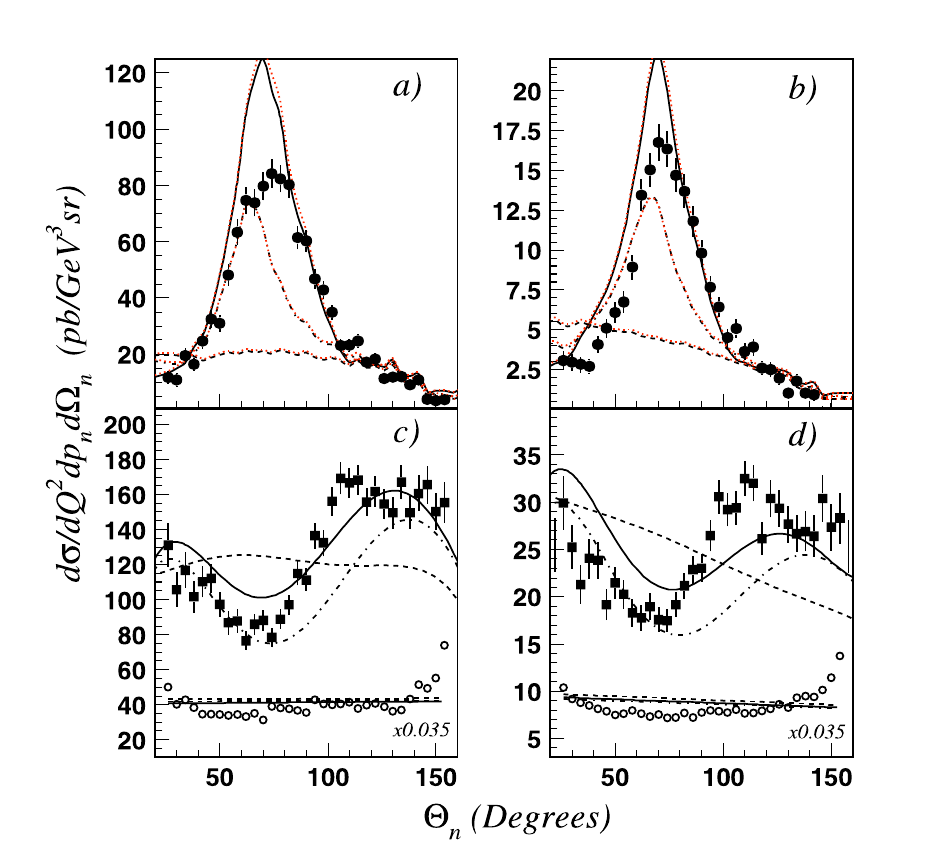}
    \caption{\footnotesize Angular distribution of recoiling neutrons measured in
      CLAS for 
      (a) $Q^2 = 2\pm 0.25$ (GeV/c)$^2$, $400 < p_n < 600 $ MeV/c,
      (b) $Q^2 = 3\pm 0.5$ (GeV/c)$^2$, $400 < p_n < 600 $ MeV/c,  
      (c) $Q^2 = 2\pm 0.25$ (GeV/c)$^2$, $200 < p_n < 300 $ MeV/c,  
      (d) $Q^2 = 3\pm 0.5$ (GeV/c)$^2$, $200 < p_n < 300 $ MeV/c. The
      data for $p_n < 100$ MeV/c are plotted in the bottom part of (c) 
      and (d), scaled by 0.035. The dashed, dashed-dotted and solid
      curves are calculations with the Paris potential using PWIA,
      PWIA+FSI and PWIA+FSI+MEC+IC respectively~\cite{Kims}.
}
    \label{CLAS_e6:1}       
  \end{center}
\end{figure}

\begin{figure}[th]
  \centering\includegraphics[scale=0.4]{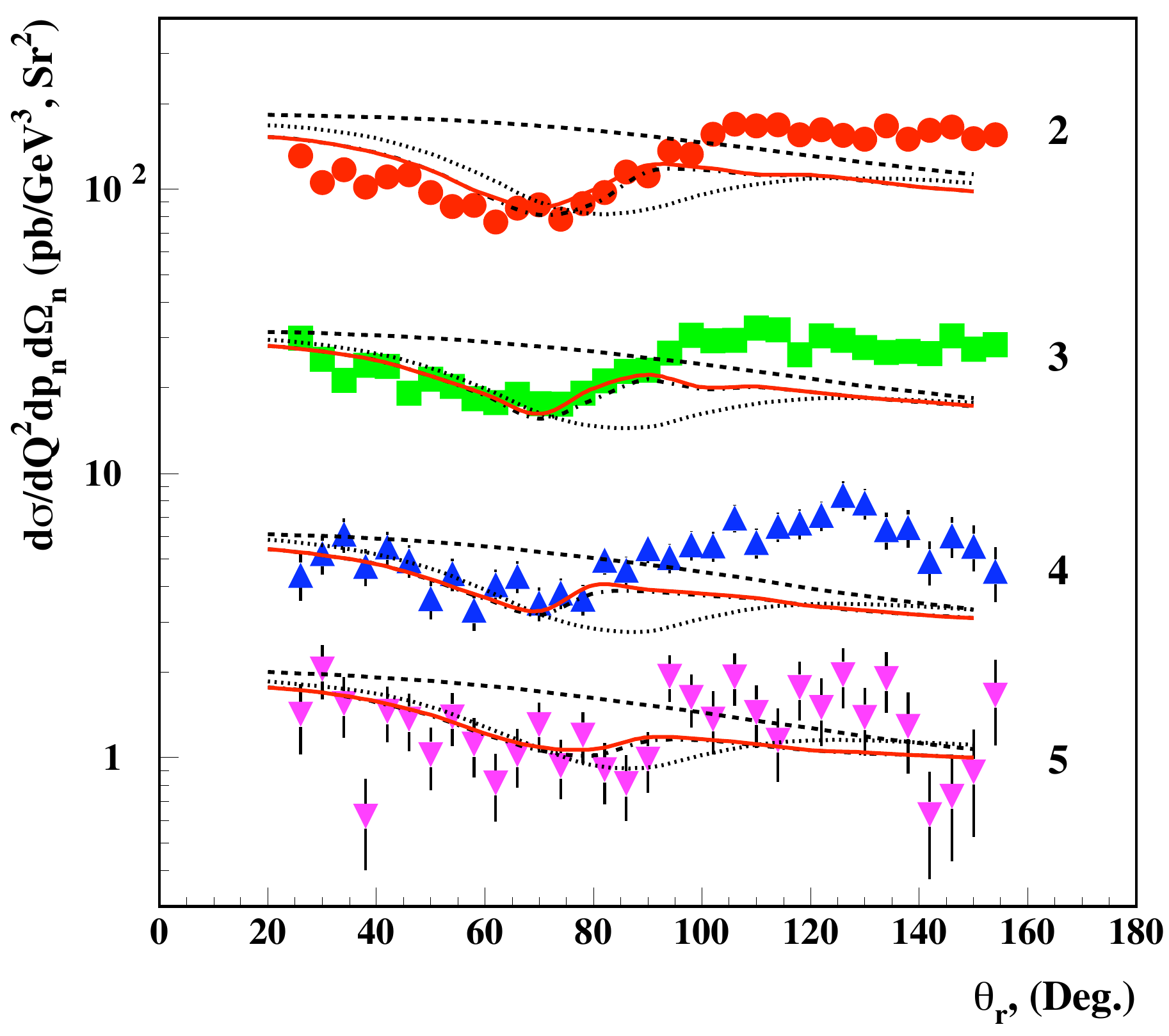}
  \caption{ \footnotesize Dependence of the differential  cross section on the direction of the recoil neutron momentum. The data are from Ref.\cite{Kims}.
Dashed line - PWIA calculation, dotted line - PWIA+ pole term of forward FSI,  dash-dotted line  -  PWIA+forward FSI,  
solid line - PWIA + forward and charge exchange FSI.  The momentum of the recoil neutron ($\ppm$)  is 
restricted to $200 < < 300$~MeV/c.   The labels 2, 3, 4 and 5 correspond to the following values of 
$Q^2 = 2\pm 0.25;  3\pm 0.5; 4\pm 0.5; 5\pm 0.5$~$\gevcsq$. No IC is included in the calculations.}
\label{Fig.ms1}
  \centering\includegraphics[scale=0.4]{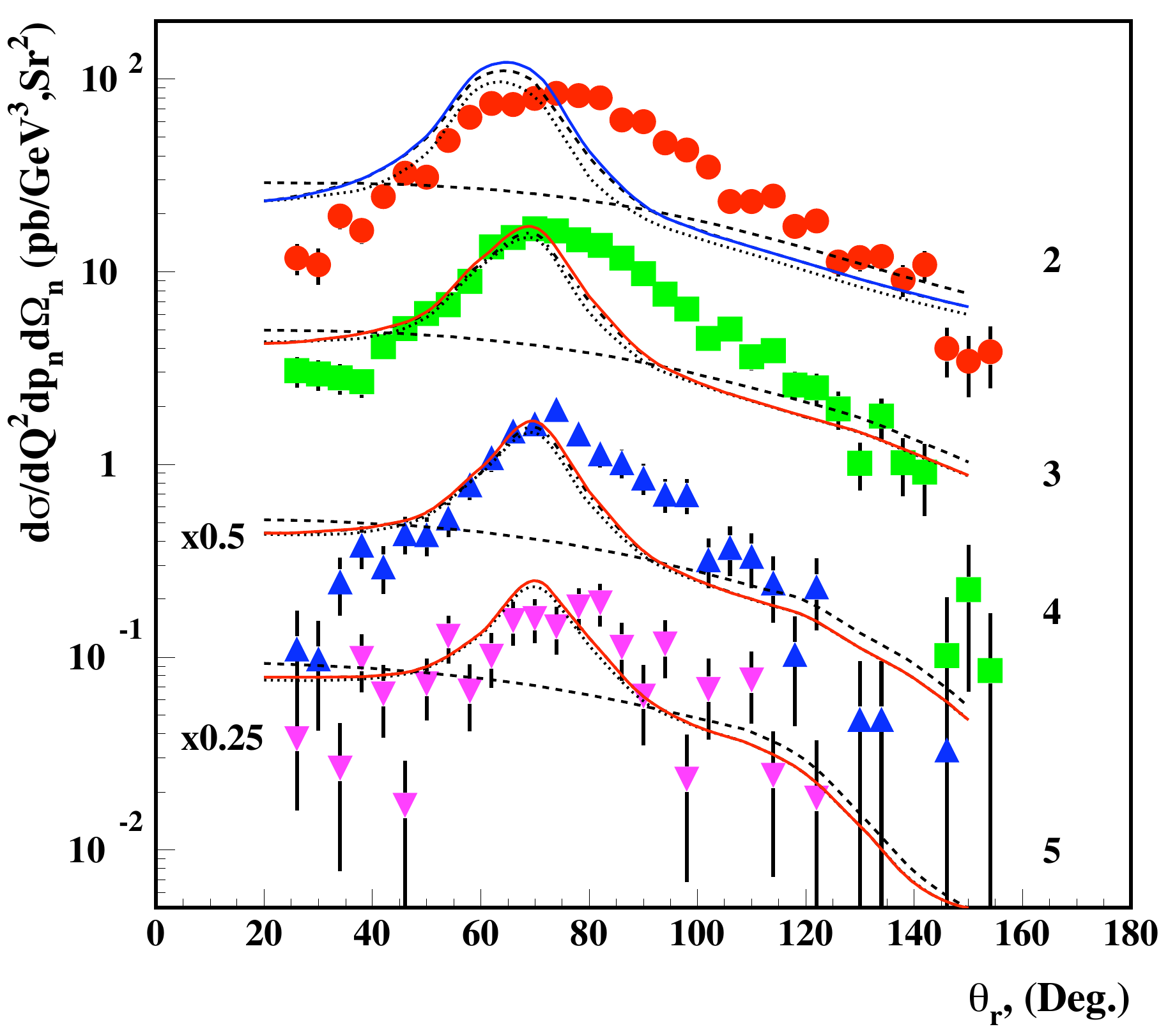}
  \caption{\footnotesize Dependence of the differential  cross section on the direction of the recoil neutron momentum. 
The data are from Ref.\cite{Kims}. Curves and labels same as Fig.~\ref{Fig.ms1}.
 The momentum of the recoil neutron is restricted to $400 < \ppm < 600$~MeV/c.   
The data sets and calculations for 
``4'' and ``5'' are multiplied by  0.5 and 0.25 respectively. 
}
\label{Fig.ms2}
\end{figure}

An important result of these experiments was that even though the final state interaction in many cases is not small it can be understood quantitatively (Figs.~\ref{CLAS_e6:1},\ref{Fig.ms1},\ref{Fig.ms2}).  Already at $\Qsq \ge 2$~$\gevcsq$ the eikonal regime is established which allows one to perform increasingly reliable estimates of these effects. Also these experiments for the first time confirmed the prediction of the eikonal approximation\cite{gea,treview} that the maximum of FSI corresponds to the recoil angle of around $70\degree$. 
This situation gives us some confidence that we can move beyond FSI to investigate the other 
characteristics of the deuteron-break up reaction. 
Note, a one-to-one relation exists between  the angle of the 
recoiling neutron ($\thetanq$) relative to the momentum 
transfer  and $\xbj$ where an increasing $\xbj$ corresponds to a decreasing $\thetanq$.

\clearpage

\subsection{Goal of the Proposal}

From the theoretical point of the view, the most intriguing question 
is how far one can extend the
boundaries of the theoretical framework based on the description of
the deuteron as a two-nucleon system?  This question can be answered
only if one starts to probe the deuteron at  extreme
kinematics corresponding to very large initial momenta of
nucleons in the deuteron.

In the CLAS experiment, cross sections for large recoil momenta have
been determined, however it was necessary to integrate over a wide
range of momentum transfers (1~$\gevcsq$) and neutron recoil angle
($0\degree - 180\degree$). As a consequence the reaction dynamics is
not well defined for these experimental cross sections and at recoil
momenta above 0.5~$\gevc$ the cross sections are completely dominated
by final state interactions.

As we will argue in this proposal, the experience we gained from the
two recent JLab experiments allows us for the first time to push our
studies to the significantly unexplored kinematic domain of probing
missing momenta up to $1$~$\gevc$ at $Q^2=4.25$~$\gevcsq$ with a kinematic
setting that is well defined, minimizes final state interactions, MEC
and IC, and suppresses the indirect reaction where the neutron is hit
and one observes the recoiling proton. 
%
%

\subsection{Proposed Measurement}
\begin{figure}[ht]
\centering
\includegraphics[width=.7\textwidth]{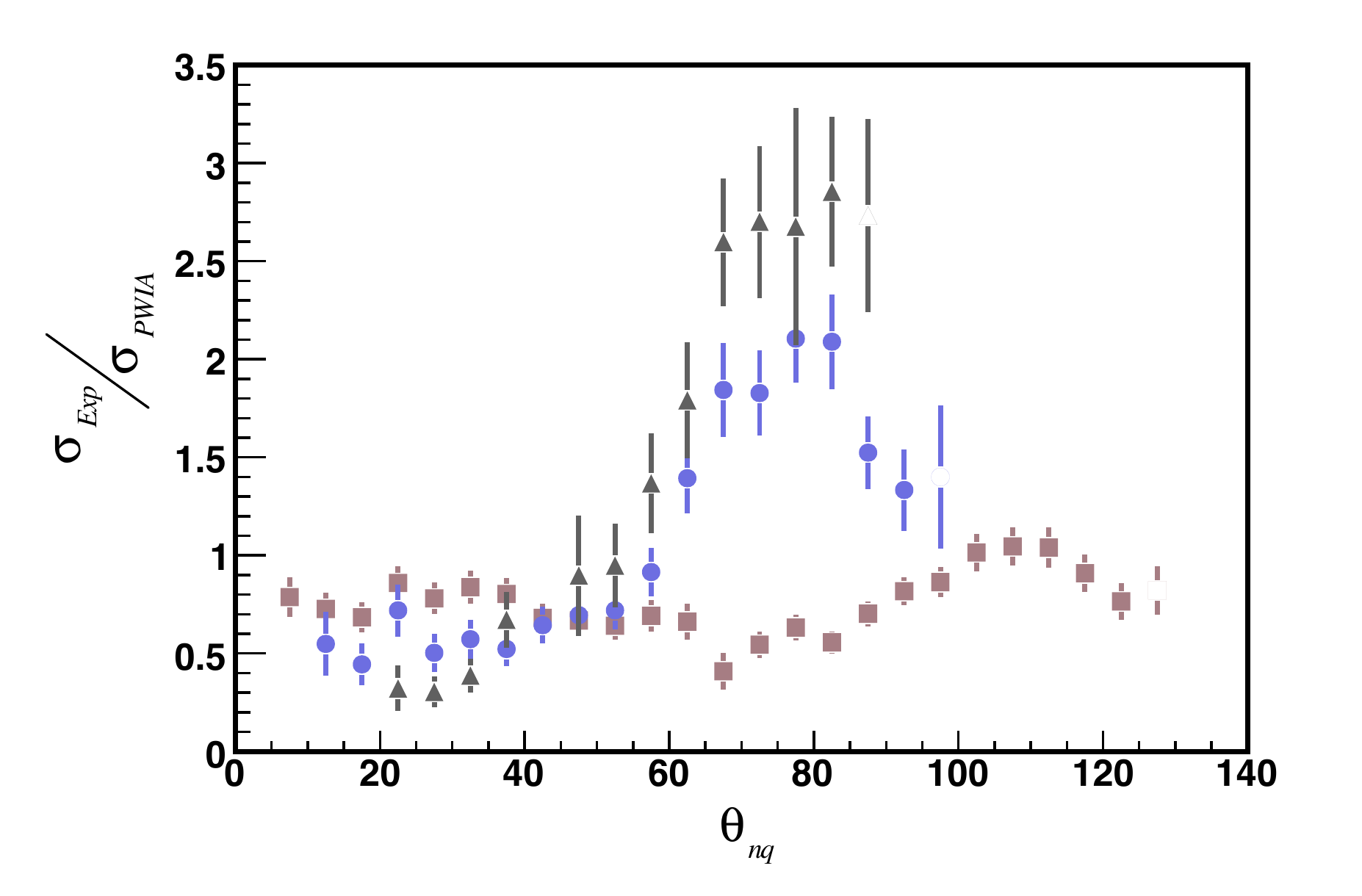}
\caption{\footnotesize The ratio $R =\sigma_{Exp}/\sigma_{PWIA}$ from the Hall A
  experiment E01-020 as a function of the recoil neutron angle $\thetanq$~\cite{LumPhd} 
for $p_m = 200$~MeV/c ( filled brown squares),$p_m = 400$~MeV/c ( filled purple circles),
$p_m = 500$~MeV/c ( filled black triangles). One can see that 
around $\thetanq = 40\degree$ the effect of FSI does not depend strongly on $\ppm$.}
\label{fige1}
\end{figure}

\begin{figure}[ht]
\centering
\includegraphics[width=.7\textwidth]{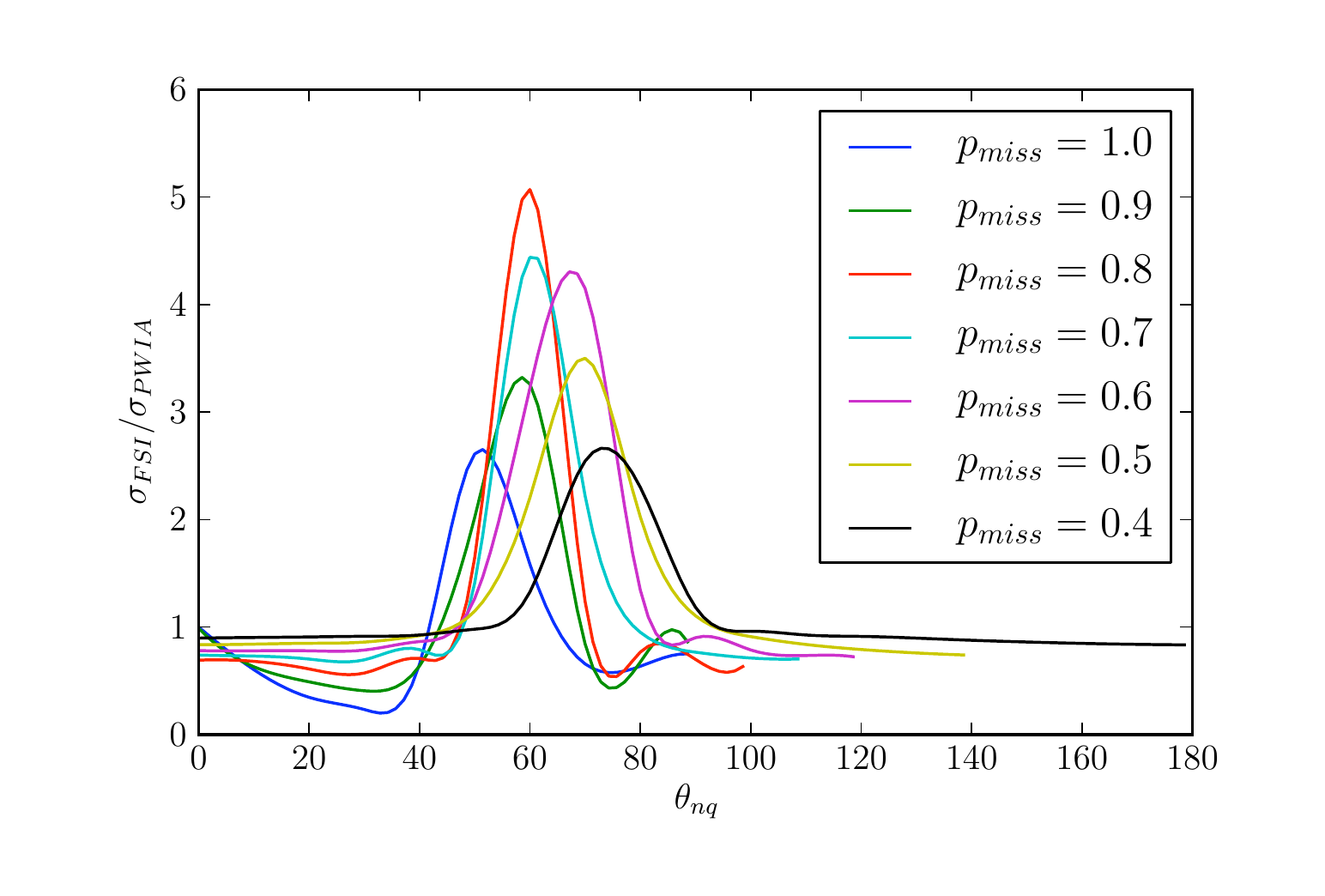}
\caption{\footnotesize The ratio $R = \sigma_{FSI}/\sigma_{PWIA}$ calculated for
  missing momenta ranging from 0.4 $\gevc$ up to 1.0 $\gevc$ and using the Paris potential. The black curve corresponds to $\ppm = 0.4$ $\gevc$ and the ratio R is maximal close to $80\degree$ while the blue curve corresponds to 1 $\gevc$ and R for this value is maximal at about $50\degree$. On can therefore see that in a region around $\thetanq = 40\degree$ the effect of FSI is only slightly dependent on $\ppm$ and within $\approx \pm 30\%$.
}
\label{fige1a}
\end{figure}

In this proposal we plan to perform an exploratory
measurement of the:
\begin{equation}
e + d \rightarrow  e' + p + n
\label{edepn}
\end{equation}
reaction probing missing momenta up to 1~$\gevc$ for one setting of
$Q^2$ and $\xbj$.  It will be for the first time that high $Q^2$
deuteron break up is probed in electro-production at such large missing
momenta and momentum transfer, at a well defined kinematic setting.
To interpret this experiment we will use three important theoretical
observations \cite{gea,treview,jeschonnek,laget,ciofi} which the previous
two experiments\cite{HallB,HallA} confirmed:
\begin{itemize}
\item Generalized eikonal approximation~(GEA) is an appropriate
  theoretical framework for the description of the reaction [\ref{edepn}] at $Q^2
  > 1$~$\gevcsq$.  These experiments confirmed that the FSI is uniquely
  defined by the missing momenta of the reaction: being dominated by
  screening effects at $p_m \le 200$~MeV/c and dominated by pure
  re-scattering effects at $p_m> 300$~MeV/c. (See Fig.~\ref{fige1}.)
\item At $p_m\ge 400$~MeV/c, the peak of re-scattering is at
  $\theta_{recoil}=70^0$ as predicted within GEA\cite{gea} and not at
  $90^0$ was was expected within conventional Glauber
  approximation. (See Fig~\ref{fige1}.)

\item The eikonal nature of FSI creates a unique angular dependence of the
  FSI effects. It is an interplay of screening~(interference of PWIA
  and FSI amplitudes) and re-scattering (square of FSI amplitude)
  effects which enter with an opposite sign in the cross section of
  the reaction.  The decrease of re-scattering effects at forward and
  backward recoil angles is associated with the increase of the
  interference effects. Since both effects are defined by the same
  re-scattering NN amplitude one arrives at approximate recoil momentum
  independence of the recoil angles at which these two effects
  significantly cancel each other. (See Fig.~\ref{fige1a}.)
 \end{itemize}
%
%
 The last point of the above observation opens a rather unexpected window
 to probe the deuteron at very high missing momenta. As it follows
 from Fig.~\ref{fige1a}, this corresponds to the recoil angles $\thetanq = \theta_r
 \approx 40 \pm 5^0$ for which FSI effects are confined within
 $\approx 30\%$ for missing momenta up to $0.95$~$\gevc$. This value of $\thetanq$ corresponds to a value of $\xbj \approx 1.35$.

\FloatBarrier

 \subsection{What can be learned from these measurements?}
 Being able to confine FSI effects within $30\%$ and pushing the
 measurements up to $p_m = 1$~$\gevc$ will allow us for the first time
 to probe the sensitivity of the scattering process to the (i)
 Reaction dynamics (ii) Deuteron wave function and (iii) Non-nucleonic
 degrees of freedom. 

\subsubsection{Reaction Dynamics}
The description of the electromagnetic interaction with bound (off-shell)
nucleons possesses many theoretical uncertainties.  The origin of the
off-shell effects in the $\gamma^*N_{bound}$ scattering amplitude is
somewhat different for low and high energy domains. In the case of low
energy transfer, the nucleons represent the quasi-particles whose
properties are modified due to the in-medium nuclear potential (see
e.g. \cite{PP}).  At high $Q^2$, the virtual photon interacts with
nucleons and the phase space volume of the process is sufficiently large. As
a result, the off-shell effects in the high energy limit are mostly related
to the non-nucleonic degrees of freedom.  Several approaches exists
to treat the off-shell effects in the high energy limit.  One of the
frequently used models is the virtual nucleon approximation (see
e.g.\cite{deFor,Sabine,treview}, in which the scattering is described
in the LAB frame of the nucleus and electrons scatter off the virtual
nucleon whose virtuality is defined by the kinematic parameters of the
spectator nucleon.  In this case the form of the wave function is
defined through the evaluation of the amplitude at the one-mass shell
pole of the spectator nucleon propagator in the Lab frame.  This
yields an off-energy-shell state of the bound nucleon.

\begin{figure}[ht]
\centering
\includegraphics[width=.7\textwidth]{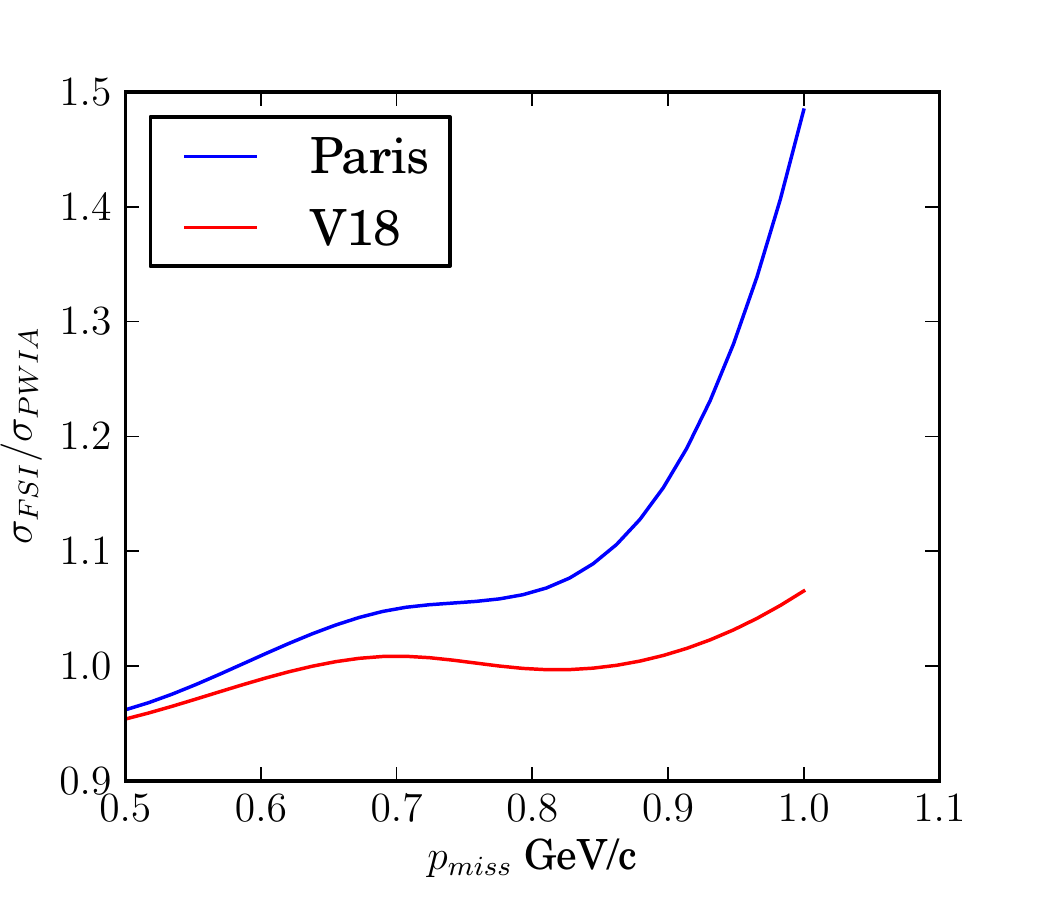}
\caption{\footnotesize
The cross section ratio  $\sigma_{FSI}/\sigma_{PWIA}$  for $\xbj \approx 1.35$ using the 
  Paris potential (blue line) and the V18 (red line), illustrating the contributions of FSI. For $\ppm \le 0.95$~$\gevc$ for both models these contributions are below 30\%}
\label{fige2}
\end{figure}

\subsubsection{Deuteron Wave Function}
Our knowledge of the deuteron wave function is restricted up to
$400~MeV/c$ relative momentum. Wave functions based on different NN
potentials start to diverge beyond this momentum range.  The uncertainty
of the deuteron wave function is not only related to the uncertainties
of the NN potential. The problem is more conceptual in a sense that
the many potentials constructed in configuration space are based on
the local (static) approximation and  become less and less relevant with the
increase of the relative momenta of the interacting nucleon. Staying
within the framework of nucleonic degrees of freedom this issues is
related to the accounting for the relativistic effects in two-nucleon
systems. These effects are significant in the region influenced by the
core of the NN interaction.

The two points discussed above are based on the nucleonic picture of both
interaction dynamics and nuclear wave function.  These approximations
have never before been applied to the large $Q^2$ kinematics when very
large missing momenta are probed.
One expects that at some point these approximations should fail
qualitatively similar to what happened in high energy large angle
photo-disintegration reactions of the
deuteron\cite{E89012,Schulte1,Schulte2,Mirazita,GG02}.  The proposed
experiment may answer at which kinematics such a breakdown occurs.

\subsubsection{Non-nucleonic degrees of freedom}
Theoretically one expects that with a recoil energy exceeding the
pion-threshold, the contributions due to non-nucleonic degrees of
freedom should become increasingly important. To date there are only
very few nuclear
experiments\cite{E89012,Schulte1,Schulte2,Mirazita,GG02} for which
such a transition is clearly observed.  These experiments played a very
significant role in the advancement of different theoretical approaches
that explicitly take into account  quark-degrees of freedom in the
nuclear interaction\cite{RNA,QGS,HRM}.
Deuteron electro-disintegration with $1$~$\gevc$ recoil momentum will be
one of such experiments.
 
We don't expect to resolve all the above issues with one such
measurement. However, this measurement will be the first in which
the kinematics are taken to the limit where a transition to non-nucleonic
degrees of freedom is expected.

\subsection{Theoretical support of these studies}
The above mentioned experiments\cite{HallB,HallA} generated significant
interest in new theoretical studies of high energy electro-disintegration
processes.  Several theoretical groups now are working on the theory of
high energy deuteron electro-disintegration
(e.g. \cite{laget_pc,sabine_pc,vanorden_pc,ciofi_pc,sargsian_pc,
  arenhoevel_pc}).  These groups established the benchmarking
collaboration to verify the agreement of their calculations at more
conventional kinematic situations\cite{bmarking}.  
Assuming that
these groups agree at low missing momentum kinematics, their
comparisons with the data at very large missing momenta will allow one to
set the limits on how much the approximations based on nucleonic
degrees of freedom can account for the cross section of the reaction.



\clearpage
\section{Experimental Program}
We plan to measure the $\deep$ cross section at kinematic settings centered
on the following missing momenta: $\ppm$ = 0.5, 0.6, 0.7, 0.8, 0.9 and
1.0 $\gevc$. Electrons will be detected in SHMS and the ejected protons in HMS.
For each setting the electron arm will remain unchanged and the
electron kinematics will be fixed at $\qsq = 4.25$ $\gevcsq$ and $\xbj
= 1.35$. 

Measurements will be done at $\ppm$ = 0.1 $\gevc$ for $\Qsq = 3.5$ $\gevcsq$ and
$\Qsq = 4.25$ $\gevcsq$. These data will be used for normalization measurements 
since at this value of $\ppm$ contributions of FSI, MEC and IC are small 
and the cross sections are large.
This has been confirmed by measurements at 
much lower $\Qsq$ values~\cite{{bl98},{NE18},{Duc92}}. 
In addition, we will also measure the $\heep$ hydrogen elastic reaction
as a cross check of spectrometer acceptance models, an additional study of target boiling effects
and a systematic check of error in beam energy, spectrometer's central  momentum and angle setting
using the kinematics of the elastic reactions.  

Fig.~\ref{fige1} shows the ratio between the experimental $\deep$
cross section determined in the E01-020 experiment~\cite{LumPhd} and
the calculated one using PWIA (using MCEEP and the PWIA model of
S. Jeschonnek). For $\ppm = 0.5$ $\gevc$ large FSI effects exist at
$\thetanq \approx 70\degree$ ($\xbj \approx 1$) as well as for
$\thetanq \leq 35\degree$ ($\xbj \geq 1.5$). For angles larger than
100$\degree$ the energy transfer is increasing ($\xbj$ is decreasing)
and one expects increasing contributions of isobar currents.

\begin{figure}[ht]
\centering
\includegraphics[width=.7\textwidth]{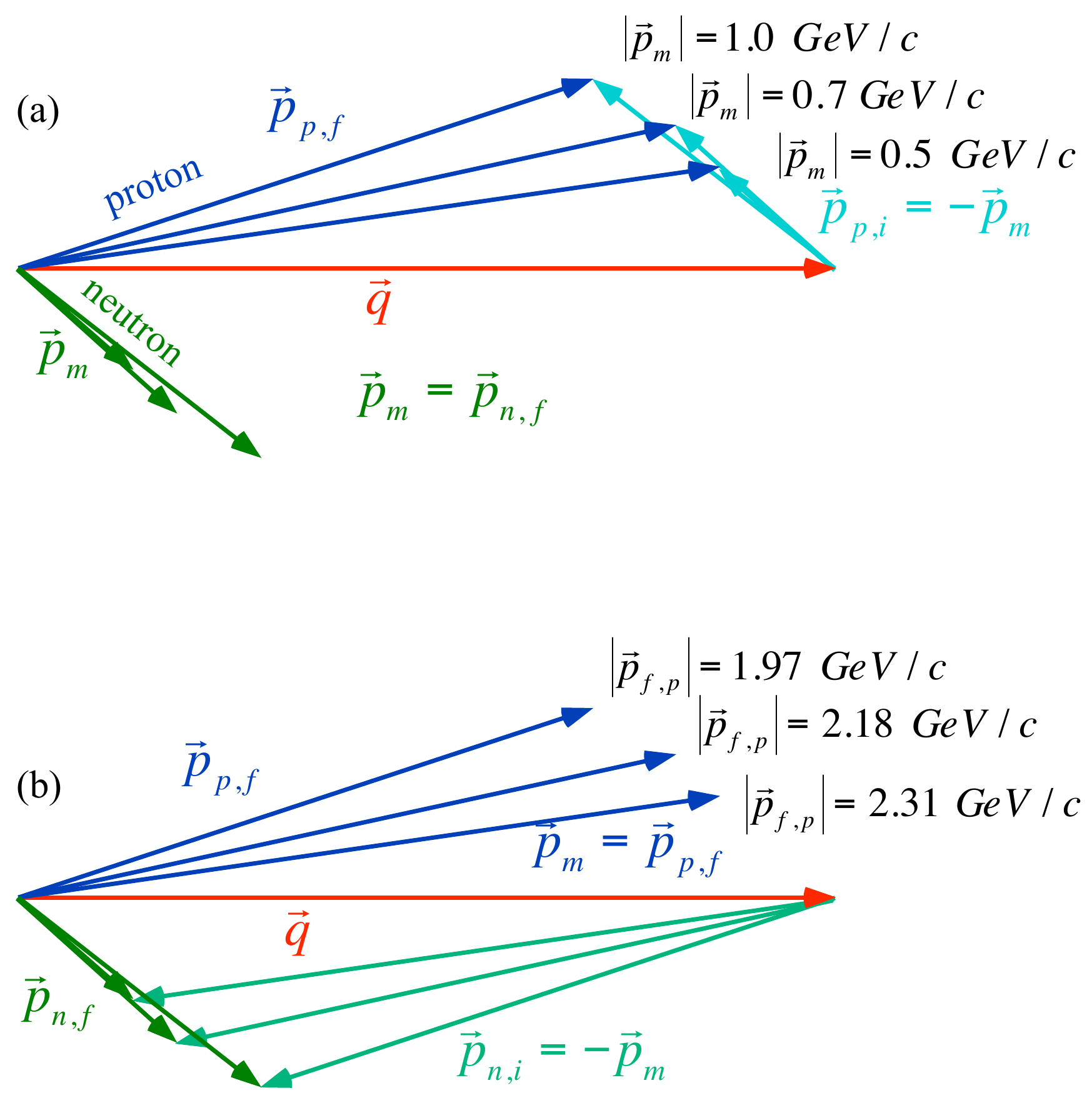}
\caption{\footnotesize The momentum vectors for the direct proton knockout (a)  and
  for the indirect reaction (b) where the proton is the spectator and
  the neutron absorbs the virtual photon.}
\label{fige3}
\end{figure}

In Fig.~\ref{fige1}, one can see that at a neutron recoil angle of
about 40 - 45$\degree$, corresponding to a value of x-Bjorken of $\xbj
\approx 1.3$, the effects of FSI are reduced to 20 - 30\% and seem to
depend only weakly on the recoil momentum. This phenomenon is
reproduced by the calculation of  M.~Sargsian
(Fig.~\ref{fige1a}). The estimated FSI effect as a function of
missing momentum for a fixed value of $\xbj \approx 1.35$ is
illustrated in Fig.~\ref{fige2} by the ratio $R =
\sigma_{FSI}/\sigma_{PWIA}$. 
It is due to this observation that we selected the
electron kinematics.
The following criteria determined the selection of the momentum
transfer:
\begin{itemize}
  \item The momentum transfer has to be large enough for GEA to be
    applicable
  \item The final proton momentum has to be significantly larger than
    the neutron recoil momentum in order to suppress the indirect
    reaction where the struck particle is the neutron and the observed
    proton is the recoiling spectator. As shown previously the
    interference of these two processes leads to a reduction of the
    cross section.
\end{itemize}

The relation between the momentum vectors for the direct and the
indirect reaction are illustrated in Fig.~\ref{fige3} 
for $\ppm = 0.5, 0.7$ and 1.0 $\gevc$. 
 Fig.~\ref{fige3}a shows the direct reaction where the proton
with an initial momentum of $\ppm = 0.7$ $\gevc$ absorbs the
virtual photon and is ejected with a final momentum of $\ppf = 2.18$
$\gevc$. For the indirect reaction (shown in  Fig.~\ref{fige3}b), a neutron 
with an initial momentum
of 2.18 $\gevc$ absorbs the photon and the recoiling proton is
observed. We expect that the probability to find a nucleon with an
initial momentum of 2.18 $\gevc$ is considerably smaller compared to
the one of finding a nucleon with an initial momentum of 0.7 
$\gevc$. Overall the ratio between the final nucleon momentum and the
recoil momentum is always larger than 1.9 in all kinematic settings and
consequently we do not expect an effect of the indirect reaction of
more than about 10\%.  The detailed kinematics can be found in
Tab.~\ref{kin_tab}.  
\begin{table}[htb]
\begin{center}
\begin{tabular}{llllllll}
$\ppm$ &  $E_f$ & $\thetae$ &  $\qmag$ &  $\ppf$ & $\thetap$ & $\thetapq$ & $\thetanq$ \\ \hline 
 0.5 & 9.322 &  11.68 & 2.658 & 2.305 &  53.47 &   8.21 &  41.19 \\ 
 0.6 & 9.322 &  11.68 & 2.658 & 2.251 &  55.60 &  10.34 &  42.31 \\ 
 0.7 & 9.322 &  11.68 & 2.658 & 2.189 &  57.63 &  12.37 &  42.06 \\ 
 0.8 & 9.322 &  11.68 & 2.658 & 2.121 &  59.61 &  14.35 &  41.07 \\ 
 0.9 & 9.322 &  11.68 & 2.658 & 2.047 &  61.56 &  16.30 &  39.67 \\ 
 1.0 & 9.322 &  11.68 & 2.658 & 1.969 &  63.49 &  18.23 &  38.02 \\ 
\hline 
\end{tabular}
\end{center}
\caption[]{Central kinematic settings for the proposed experiment. The incident energy assumed is $\einc = 11.0\,\gev$. The electron kinematics is held fixed at $\xbj = 1.35$ and $\qsq = 4.25\gevcsq$. }\label{kin_tab}
\end{table}

\begin{figure}[Ht]
\centering
\includegraphics[height=.35\textheight]{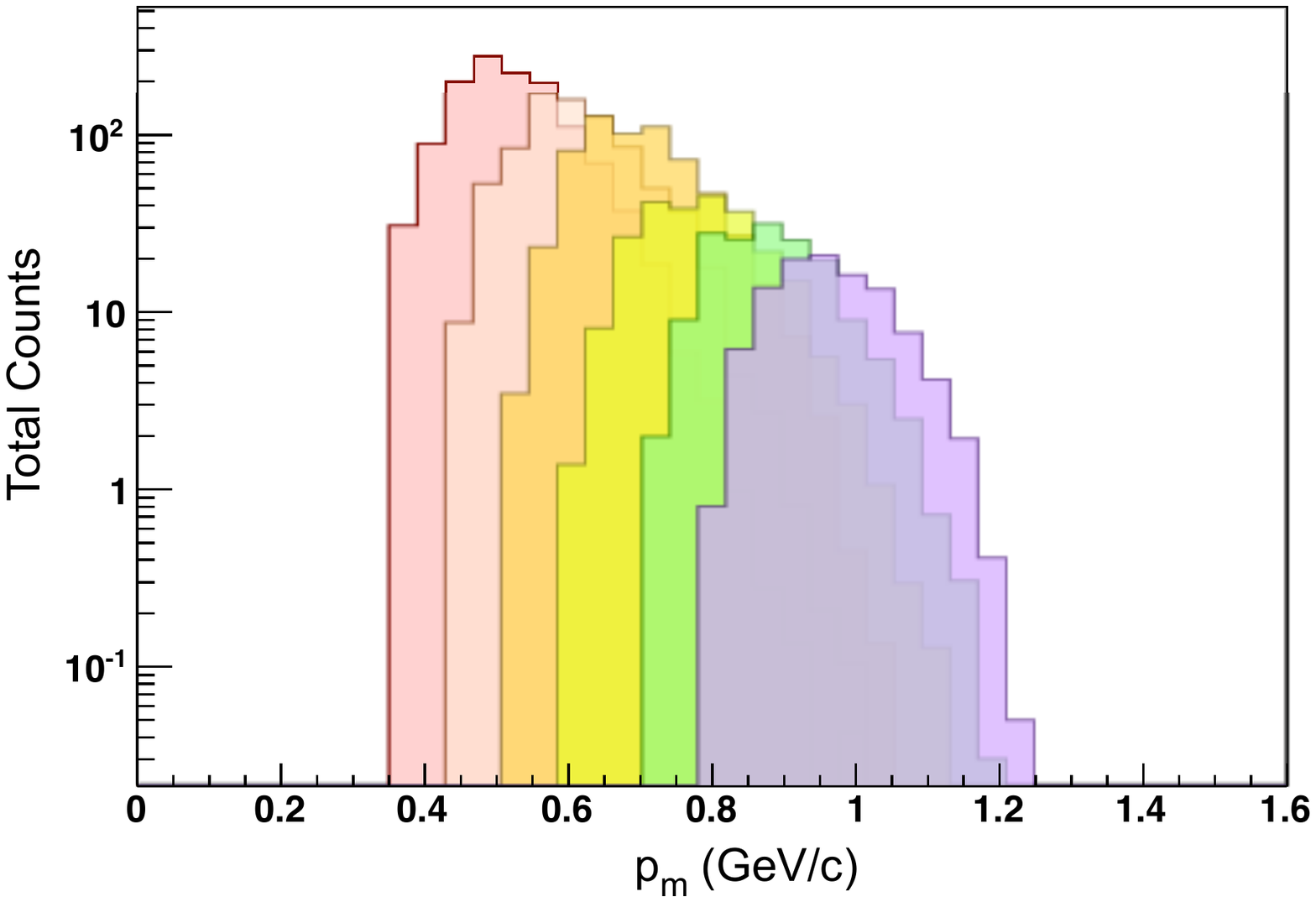}
\caption{\footnotesize Acceptance in missing momentum for the proposed kinematic 
  settings. The cuts described in the section on count rates have
  been applied.}
\label{fige4}
\end{figure}

The acceptance in missing momentum for each kinematic setting is shown
in Fig.~\ref{fige4}. 
\FloatBarrier
Clearly the different setting have considerable
overlap. We plan to use this overlap to obtain a continuous data set
of cross sections between a missing momentum of 0.5 and 1.0 $\gevc$. The
estimated statistical errors of the data are indicated in
Fig.~\ref{figc2}. We expect that this experiment is dominated by the
statistical error since one typically obtains a systematic error of
the order of 5\%. The expected statistical errors range from 5\%
for the lower missing momenta to 20\% for 1.0 $\gevc$. Given that this
kinematic region can be considered as an unexplored new territory we
believe that a 20\% measurement is still very valuable.
%
%
%


\FloatBarrier

\clearpage
\section{Count-Rates}
\begin{figure}[ht]
\centering
\includegraphics[width=.65\textwidth]{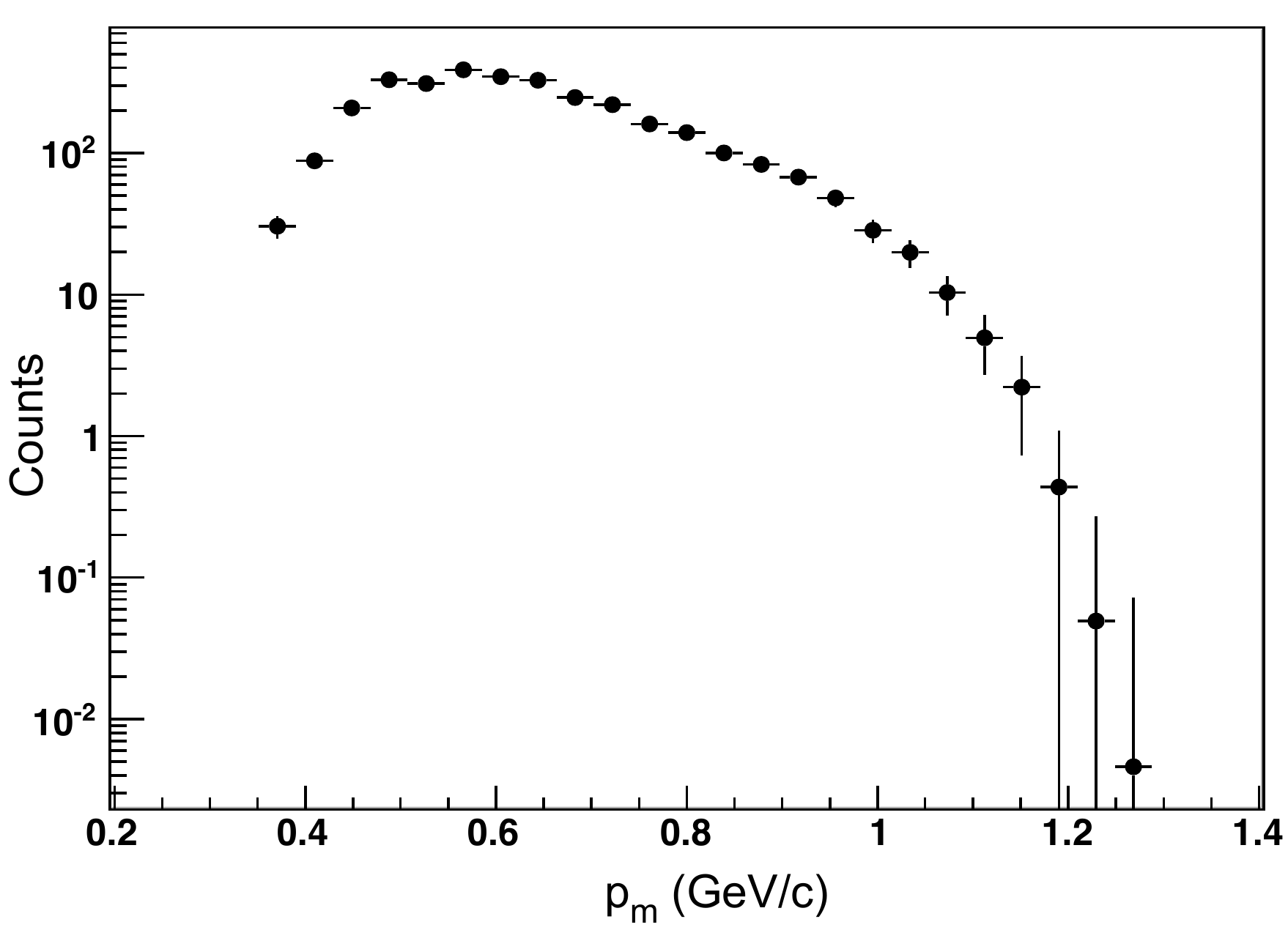}
\caption{\footnotesize Total counts expected per missing energy bin. Included are all
  the cuts described in this section and overlapping kinematic
  settings have been added.}
\label{figc1}
\end{figure}

\begin{figure}[ht]
\centering
\includegraphics[width=.7\textwidth]{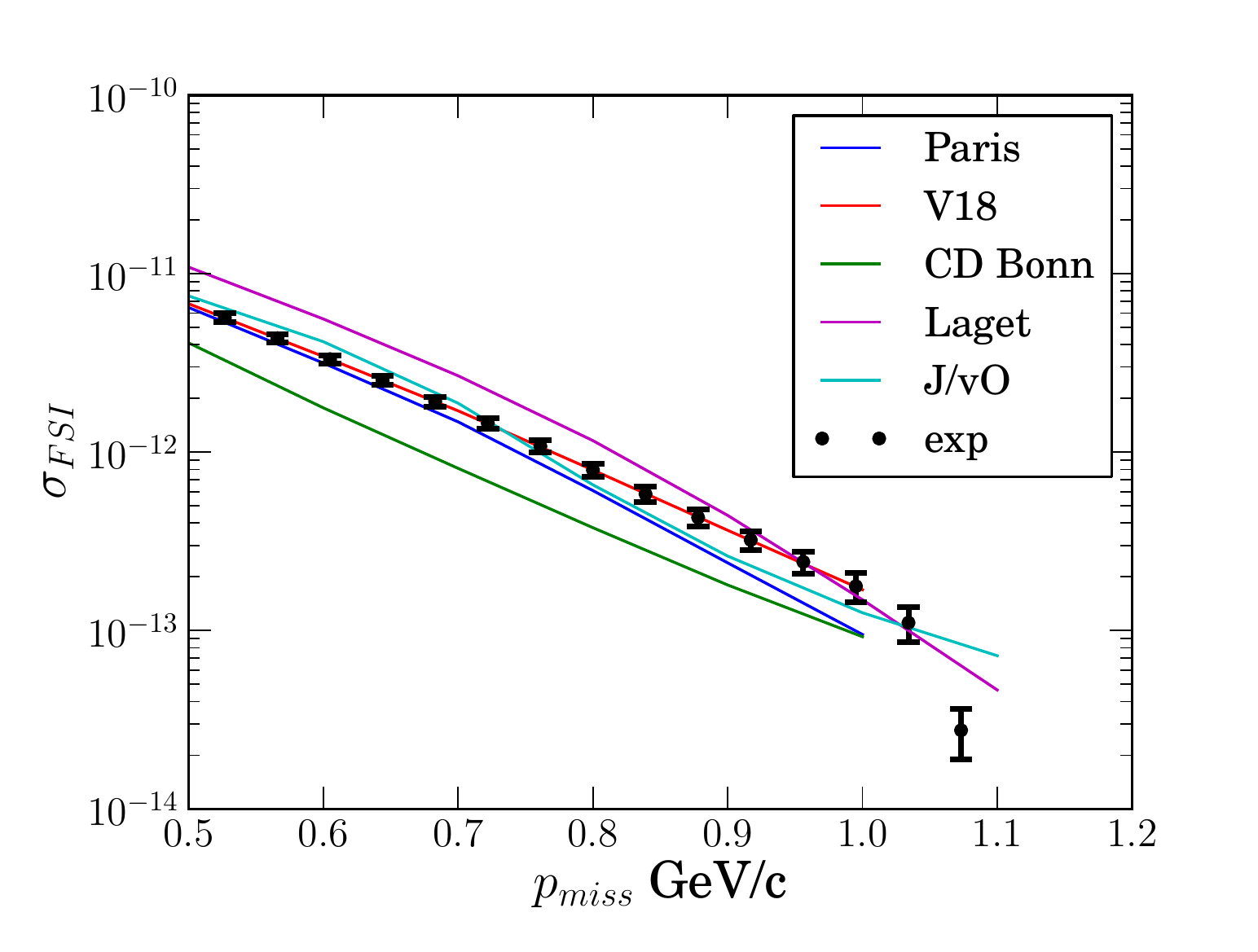}
\caption{\footnotesize The expected statistical error as a function of missing momentum compared to a range of calculations including FSI and different models for the deuteron wave function.}
\label{figc2}
\end{figure}

The coincidence count-rates for electrons in SHMS and protons in HMS have been estimated using the Hall-C
monte-carlo program SIMC~\cite{SIMC}. The coincidence cross section
has been calculated within the PWIA using the V18 momentum distribution and included radiative effects. The following cuts
have been applied for the rate estimates:
\begin{description}
   \item[electron solid angle : ]
     \mbox{$-0.05 \le \thetae \le 0.05$,    $-0.025 \le \phie \le 0.025$},
     \\ angles are in radians
   \item[electron momentum acceptance :]
     $-0.08 \le \Delta p/p \le 0.04$ 
   \item[proton solid angle: ]
     \mbox{$-0.06 \le \thetap \le 0.06$, $ -0.035 \le \phip \le 0.035$}
   \item[proton momentum acceptance :]
     $-0.1 \le \Delta p/p \le 0.1$ 
   \item[Bjorken-x:] $ 1.3\le \xbj \le 1.4$
   \item[missing momentum :] missing momentum bin width $=\pm 0.02$ $\gevc$
   \item[missing energy :] $-10\leq \emiss \leq 25$ $\mev$
   \item[momentum transfer :] $\Qsq = 4.25 \pm 0.25 $ $\gevcsq$
\end{description}
A 15 cm liquid deuterium target and a current of 80$\mu$A have been
assumed, which results in a luminosity of $L =3.2\exx{38}$ $\rm
cm^2\cdot sec^{-1}$. The results of these estimates are shown in
Fig.~\ref{figc1} in which the counts per hour (after combining different HMS settings)
 are plotted as a function of missing momentum. In Fig.~\ref{figc2}, the estimated
statistical errors  are compared to calculated 
cross sections using different models 
for the deuteron wave function and final state interactions.

\begin{figure}[ht]
\centering
\includegraphics[width=.6\textwidth]{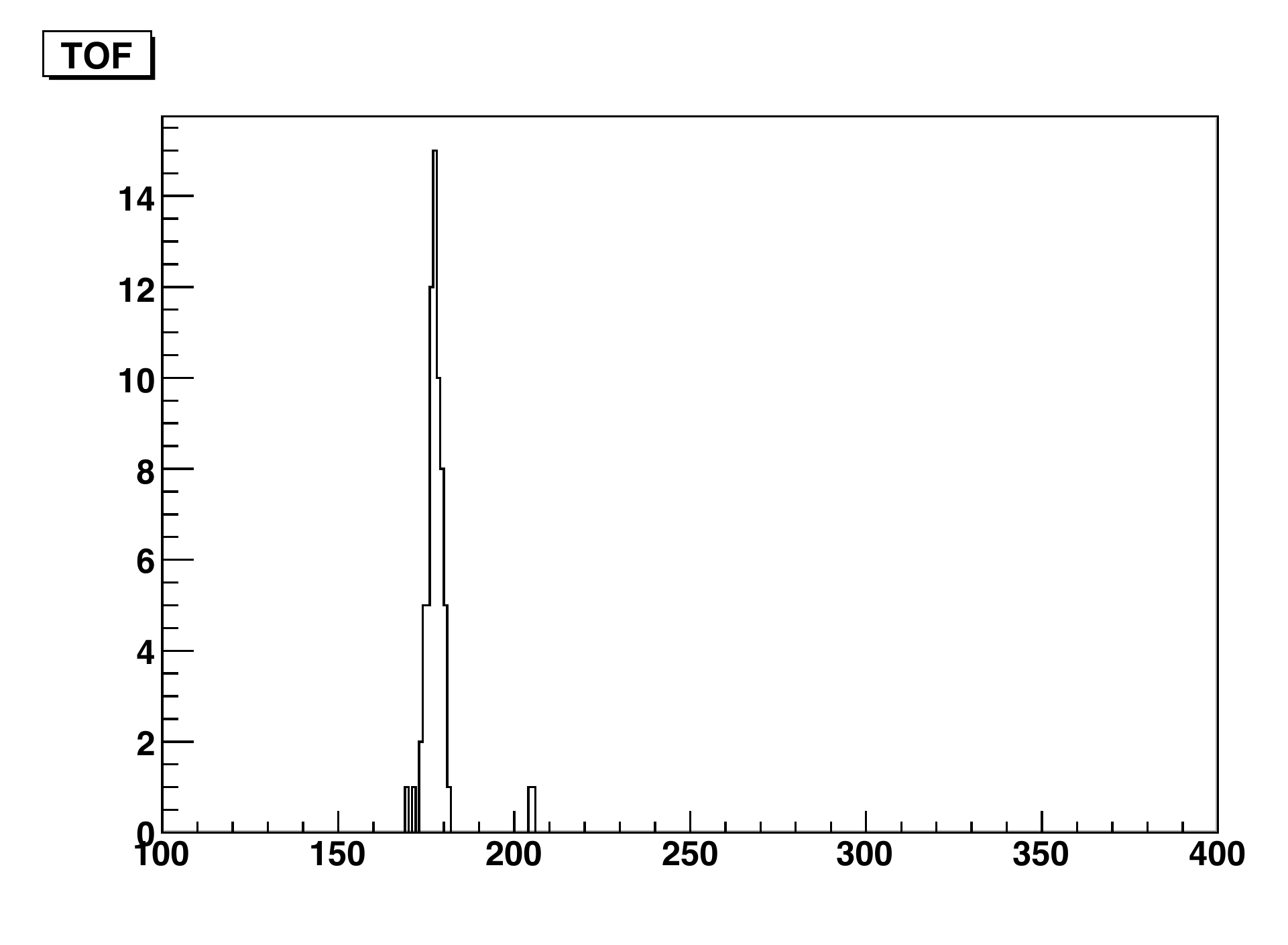}
\caption{\footnotesize Time of flight spectrum between the two spectrometers as
  obtained in the E01-020 experiment for $\qsq = 3.5$ $\gevcsq$,
  $\ppm = 0.5$ $\gevc$ and $\xbj \approx 1.45$.}
\label{figc3}
\end{figure}

\FloatBarrier
The proposed electron kinematics differs from the one of the 
previous $\deep$ experiment (E01-020) in that the momentum transfer 
has increased from 3.5 to 4.25 $\gevcsq$ and the incident energy has been doubled. The singles rates measured previously in E01-020 were about 1 KHz for electrons 
and about 400 Hz for protons for the $\ppm = 0.5$ $\gevc$ setting. The inclusive electron scattering code INCLUSIVE by M.Sargsian~\cite{INCL}, which reproduces inclusive cross sections quite well, has been used to estimate the electron singles rates for the new kinematics. The results showed that the rates are very similar to those of the Hall~A experiment.
We therefore expect similar electron singles rates and similar or lower proton rates for the proposed kinematics. No corrections for accidental coincidences were necessary in the
analysis of the $\qsq = 3.5$ $\gevcsq$ kinematics of E01-020 data, due
to the small single rates (Fig.~\ref{figc3}).

Using the code EPC to estimate the variation of the proton singles
rate at the spectrometer settings for the higher missing momenta
measurements, we found that it is expected to increase by a factor of 1.6
for the highest missing momentum setting. At this setting the overall
signal to noise ratio, using the full acceptance of the spectrometers,
a timing window of 2.5 ns, and without a cut in missing energy was
estimated to be 1:1 and we expect this ratio to be much higher once
all cuts have been applied. 

Proton and electron singles rates are well within the capabilities of the
spectrometer detector systems. The resulting signal to noise ratio is 
generally large and we do not anticipate any background problems.
In E01-020 we found the pion rates to be generally well below the
singles rates for electron and protons. In the electron arm, pions
will be rejected by using the calorimeter and the noble gas 
Cherenkov detector. 
In the hadron arm, pions can be rejected using
time-of-flight measurements since the momenta involved are below 2.3
GeV/c and the corresponding time-of-flight difference between pions
and protons is $\ge 5.8$~ns. 
In addition pion events produce a
continuous missing energy spectrum and no significant pion background
has been found in the previous experiment.

\clearpage
\section{Beam Time Request}
We plan to measure a total of 9 different kinematic settings
(including the hydrogen elastic calibrations). Table~\ref{bt1} shows
the summary of the requested beam time. The beam time on target
required to achieve the necessary statistics includes the following
items:
\begin{itemize}
	\item Time to determine the spectrometer pointing at each setting 
	\item Time for target and spectrometer changes
\end{itemize}
The two low $\ppm$ measurements are calibration measurements that
overlap with the Hall A experiment.

\begin{table}[htb]
\begin{center}
\begin{tabular}{lccc}
  $\ppm$ $\gevc$ & Data Taking  & Overhead & Sub-total \\ \hline
 0.1 ($\Qsq = 3.5$ $\gevcsq$) &4      & 2.0    &6     \\
 0.1  &4      & 2.0    &6     \\
 0.5  &41     & 2.0    &43   \\ \hline
 0.6  &47     & 2.0    &49   \\
 0.7  &67     & 2.0    &69   \\
 0.8  &60     & 2.0    &62   \\ \hline
 0.9  &82     & 2.0    &84   \\
 1.0  &137    & 2.0    &139  \\ \hline
        &             &         &       \\
Optics Commissioning &       &         & 16    \\
Target Commissioning &       &         & 16    \\ \hline
$\heep$ calibrations & 2.0   & 4.0   & 6.0  \\
               &             &         &       \\
TOTAL          &             &         & 496\\ \hline \hline 
\hline
\end{tabular}
\end{center}
\caption[]{Beam Time Overview}
\label{bt1}
\end{table}
\FloatBarrier


\end{document}